# The effect of cerium doping and rate of crystallization on the microstructure of the Al$_2$O$_3$/YAG eutectic composite

L.O. Hryn, Yu.V. Siryk, V.V. Baranov, O.M. Vovk, S.V. Naydenov, S.V. Nizhankovsky

Institute for Single Crystals of the National Academy of Sciences of Ukraine

Abstract

Al$_2$O$_3$/YAG and Al$_2$O$_3$/YAG:Ce$^{3+}$ eutectics (up to 1 at.%) with a Chinese script microstructure were obtained by the method of horizontal directional crystallization (HDC) in a carbon-containing reducing atmosphere based on Ar. The effects of the pooling rate and Ce$^{3+}$ dopant concentration on the microstructure of the eutectic were studied by the modified chord method. The deviation of the dependence of the characteristic eutectic distance ($\lambda_{eut}$) on the pooling rate ($V$) from the Jackson-Hunt model for the Al$_2$O$_3$/YAG eutectic has been established. It was shown that in the range $V$ = 5-50 mm/h, the relation $\lambda_{eut}^{1.75} \times V = const$ was satisfied. A change in the pathway for restructuring the structure of the doped eutectic at a certain critical pooling rate was revealed, which may indicate that the development of morphological instability of the crystallization front is carried out as a two-stage process. The conditions for obtaining doped eutectic Al$_2$O$_3$/YAG:Ce$^{3+}$ of high homogeneity with a minimal characteristic eutectic distance $\lambda_{eut}$ and a minimal difference between the parameters of coarse and fine microstructure have been established. It was shown that increasing cerium ion concentration causes an increase in the structure parameters and its value deviation. Intensive formation of the third phase occurred when the limit of cerium ion concentration was exceeded.

**1. Introduction**

White light-emitting diodes (WLEDs) are becoming more popular than conventional light sources due to their longer lifespan, high luminous efficiency, environmental friendliness, and cost-effectiveness. One of the most common methods for manufacturing WLEDs is the combination of blue InGaN LEDs with a luminescent converter (LC) made from yttrium aluminum garnet phosphor powder doped with cerium ions Y$_3$Al$_5$O$_{12}$:Ce$^{3+}$ (YAG:Ce$^{3+}$) and an epoxy resin compound [1]. However, the thermal instability of powder composite phosphors with an organic component shortens the service life and worsens the color characteristics of powerful WLEDs [2]. To solve these problems, researchers have recently considered a crystalline eutectic composite of Al$_2$O$_3$/YAG:Ce$^{3+}$ as a material for LC, obtained using various directional crystallization methods, particularly the Bridgman, Czochralski, floating zone, and horizontal directional crystallization (HDS) methods. and micropooling [3-8]. Due to its high heat resistance (up to 1973 K), stable luminescent characteristics, high mechanical properties, and excellent luminous efficiency, this material is one of the most promising for solving these problems [9,10]. Among the methods for producing Al$_2$O$_3$/YAG:Ce eutectic, the HDS method stands out due to the possibility of obtaining ingots of sufficiently large size [11] and controlling the distribution of matrix components and impurities in the ingot, which affect the functional characteristics of the material. When $Al_2O_3$/YAG eutectic is obtained by directional crystallization methods, a 3D interpenetrating microstructure is formed [12], which in the literature has acquired the name "Chinese script" [13]. In addition, when obtaining a two-phase eutectic $Al_2O_3$/YAG:$Ce^{3+}$, as a result of concentration supercooling, a colony structure is formed, and the microstructure is distributed into regions with smaller (fine structure) and larger (coarse structure) eutectic distances [14]. This leads to a redistribution of the areas of structural elements of eutectics, which changes the homogeneity of the microstructure. This irregular type of microstructure differs from traditional regular microstructures of the lamellar or fibrous type by a significant scatter in the sizes of domains of individual phases that affects the calculation of the so-called *eutectic distance*.

According to the Jackson-Hunt model [15], which was introduced for regular eutectics, the eutectic distance ($\lambda$) obeys the formula $\lambda^2 \times V = const$, where $V$ is the solidification rate of the eutectic. The authors of [16] showed that for irregular eutectic systems of the $Al_2O_3$/YAG type, a deviation from this model is observed.

When characterizing the microstructure of an irregular eutectic, the value $\lambda$ is called the phase interval [17], gap [18], interphase length [12], and average eutectic interval [19]. In our opinion, for an irregular eutectic, a more correct name for the parameter $\lambda$ is the characteristic eutectic distance ($\lambda_{eut}$).

When using a eutectic composite as a converter for WLEDs, the stability of the lighting characteristics of the luminous flux over the area of the LC is important. The transverse size of the structural elements of the $Al_2O_3$ and YAG eutectic phases (i.e., the parameter $\lambda_{eut}$) and the concentration of cerium affect these lighting characteristics; their stability is ensured by greater homogeneity of the structural elements of the microstructure and the distribution of cerium throughout the volume of the LC.

This work continues a series of studies on $Al_2O_3$/YAG eutectic doped with $Ce^{3+}$ ions for use as converters for high-power white light sources [20-22].

The purpose of this work is to study the features of the formation of coarse and fine microstructures of the $Al_2O_3$/YAG:$Ce^{3+}$ eutectic and to establish the patterns of their distribution.

**2. Experiment**

2.1. Preparation of $Al_2O_3$/YAG and $Al_2O_3$/YAG:$Ce^{3+}$ eutectics

For homogenization, high-purity oxide powders $Al_2O_3$ (99.999%) and $Y_2O_3$ (99.99%) were mixed in the appropriate molar ratio $Al_2O_3$/$Y_2O_3$ = 81.5/18.5, which corresponds to the eutectic composition. Ce ion dopant was added in the form of $CeO_2$ powder (99.9%) with Ce atom concentrations of 0.25 at.%, 0.5 at.% and 1 at.% relative to Y. The $Y_2O_3$ content was corrected according to the substitution of $Y^{3+}$ ions for $Ce^{3+}$ ions in the YAG crystal lattice.

The mixture was subjected to uniaxial compression at 40 MPa to obtain tablets with a diameter of approximately 30 mm and a thickness of 10 mm, which were then sintered at 1573 K for 12 hours in air. The sintered tablets were placed in a boat-shaped molybdenum crucible. The process of producing eutectic ceramics $Al_2O_3$/YAG and $Al_2O_3$/YAG:$Ce^{3+}$ was carried out in a modified HDS Gorizont-3 furnace at an Ar (99.993%) pressure of 0.1 MPa. The tablets were melted with a melt superheat of 50-95 K above the solidification temperature of the eutectic and kept for 1 hour. Then, the crucibles were pulled at different speeds through the gradient zone. In this work, the pulling speed ($V$) was 5 mm/h, 15 mm/h, 30 mm/h, and 50 mm/h. The crystallization rate of the ingot regions from which samples were cut for research approximately corresponded to the rate of pulling the crucible. Details of the experiment are described in [23]. The temperature gradient ($G$) in the melt crystallization zone was ~ 45 K/cm. The choice of temperature gradient was determined as follows: at a low-temperature gradient, the processes of atomic diffusion and mixing in the melt at the boundary of the phase distribution slow down significantly due to the significant viscosity of the melt [24], while exceeding this value can lead to too high overheating of the melt (>100 K) inside the hot zone, which depends on the design features of the thermal zone and will lead to the formation of an accompanying metastable $YAlO_3$ perovskite phase according to the phase diagram [25].

$Al2O3$/YAG and $Al_2O_3$/YAG:$Ce^{3+}$ eutectic ingots with dimensions of 60 mm × 30 mm × 15 mm were obtained; samples for research measuring 7 mm × 7 mm × 0.5 mm were cut from the central part of the ingots, where stabilized conditions for the crystallization process were achieved. Samples for research underwent finish polishing with ASM 3-2 diamond powder.

The phase composition of the eutectic was determined by analysing powdered samples of the eutectic using a DRON-3 X-ray diffractometer. The microstructure was observed using a scanning electron microscope (SEM) JSM-6390LV (JEOL Ltd., Japan). The study of the morphology and determination of the surface roughness of the samples was carried out using a scanning probe microscope Solver P47H-PRO (NT-MDT).

2.2. A modified chord method for analysing the structure of eutectics

The use of known methods for determining $\lambda_{eut}$ and other structural parameters, which are applied to eutectic composites with a regular microstructure, leads to large measurement errors in the case of irregular microstructures [24]. As a result, it is difficult to establish the real dependence of the obtained microstructure on the eutectic growth conditions. In this work, to estimate $\lambda_{eut}$ and obtain various other information about the microstructure of $Al_2O_3$/YAG:$Ce^{3+}$, the method of constructing secant lines was used [26]. The characteristic dimensions of the eutectic structure were obtained from the statistical analysis of the lengths of the chords cut off by the secant line in the dark ($Al_2O_3$ phase) or light (YAG phase) areas in the image of the eutectic section. When crossing the image, the line forms a series of alternating dark and light chords (Fig. 1a). The eutectic distance $\lambda_{eut}$ was taken to be the length of the binary domain "dark chord ($Al_2O_3$) - light chord (YAG)", which occurs most often in accordance with the statistical distribution (Fig. 1b). In addition, for the analysis of this structure, binary domains of the maximum length $\lambda_{max}$, as well as the lengths of light and dark chords, were considered separately (averaged eutectic intervals $\lambda_{eut}^{Al_2O_3}$ and $\lambda_{eut}^{YAG}$).

A special computer program for image processing was created and used to study images of eutectic slices using the method of secant lines. Image processing consisted of preprocessing (removal of noise, brightness gradients), scanning of the image with parallel secant lines at a given angle, with a distance of 1-4 μm between them (on the scale of the structure under study), after which the total array of chord lengths obtained was calculated along all secant lines.

To eliminate the effect of texture, a series of scans was first performed for each sample, in which the scanning direction was sequentially changed along all directions of the SEM image with a step of 3°. After this, the direction along which the domain sizes were maximum was determined, and the value of $\lambda_{eut}$ was calculated in the direction perpendicular to the defined direction.

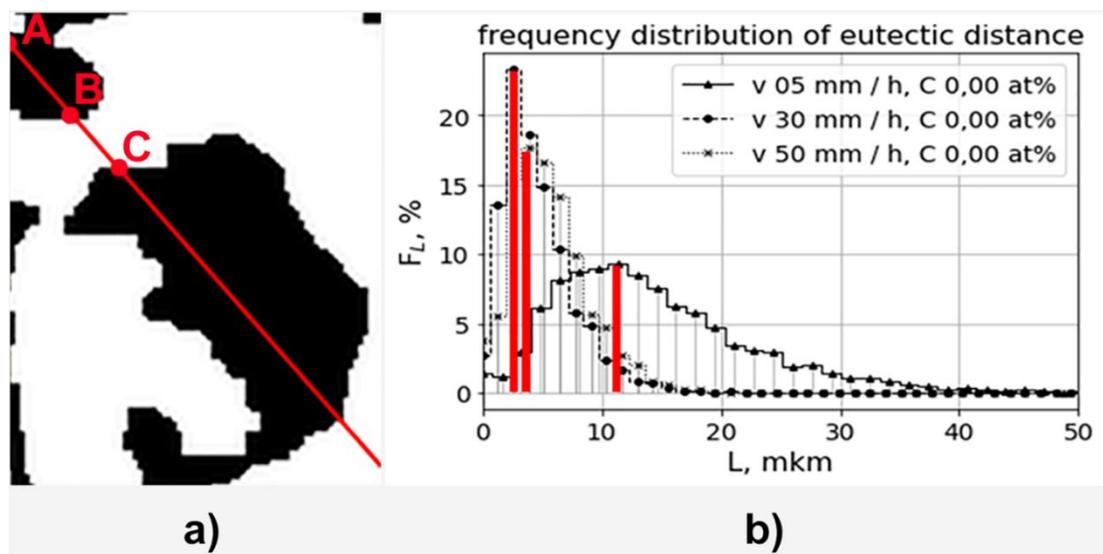

**Fig. 1**. *a)* a fragment of a binarized image and a secant line: segment AB – black chord ($Al_2O_3$ phase), BC (YAG phase) – white chord, AC – binary domain; *b)* examples of the obtained frequency distributions of binary domains by their lengths for three different samples. $F_L$ is the percentage of chords falling into each interval.

The amplitude and periodicity of the alternation of areas of local increase or decrease in the size of the domains, the so-called coarse and fine structure along the chosen direction, were studied using the method of constructing averaged profilograms. This method is as follows. First, a translational series of secant lines of the same length, with a step of 2-5 μm, is built at a given angle of inclination. After that, each secant line is divided into equal intervals (channels) in such a way that each channel must coincide with similar channels of other lines of the translational series. For each domain on the section, its position (the distance of the domain from the beginning of the secant line) is determined, as well as the number and position of the channel in which it falls. Next, you need to sum up the values of the domain lengths in all matching channels of the translational series and obtain the average value of the domain length Λ using the formula (1):

$$\Lambda_i(M) \equiv \frac{1}{M}\sum_{m=1}^{M} L_i^{(m)}, i = 1, \ldots, N \qquad (1)$$

where $N$ is the number of channels, $L$ is the length of the domain falling into the channel with the number $i$, and $M$ is the number of secant lines.

The obtained averaged profilogram is a diagram on which the channel numbers or their positions on the secant line are plotted along the $X$-axis, and the values of Λ for each channel are plotted along the $Y$-axis. An example of the constructed averaged profilogram is shown in Fig. 5 c.

## 3. Results and discussion

The phase composition for all ingots of the obtained eutectic ceramics was the same. Only two phases, $Al_2O_3$ and $Y_3Al_5O_{12}$, were observed in the diffraction patterns. Examples of the obtained diffraction patterns are shown in Fig. 2.

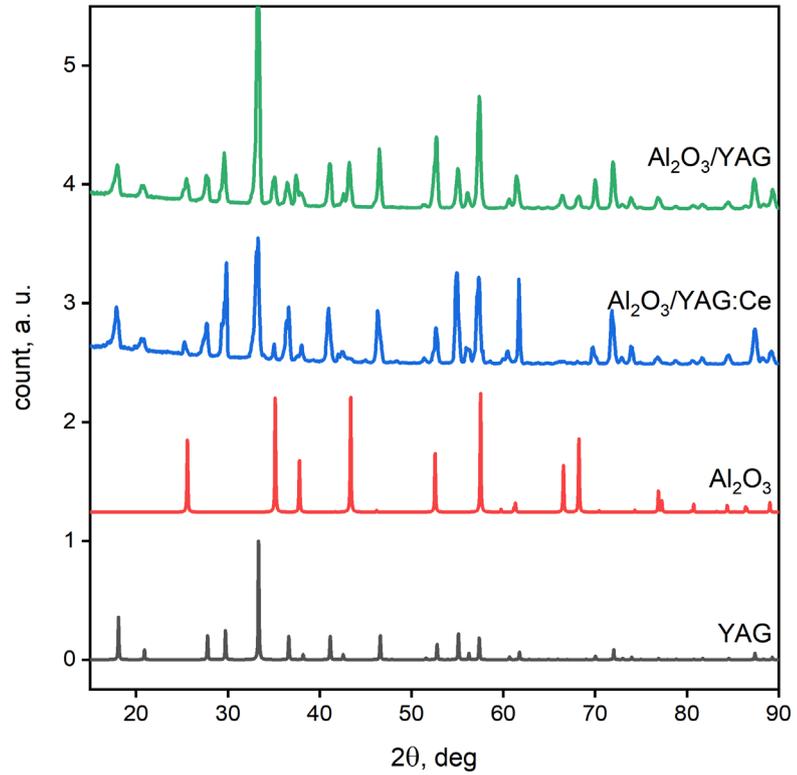

Fig. 2. Diffraction patterns of $Al_2O_3$/YAG and $Al_2O_3$/YAG:Ce (0.25 at. %) eutectic samples obtained by the HDC method at $V = 30$ mm/h and their comparison with $Al_2O_3$ and YAG samples.

The diffraction patterns did not reveal reflections of the crystalline phase of perovskite $YAlO_3$, which can be formed when the melt is overheated above a certain temperature [21]. Determination of the phase containing Ce is beyond the sensitivity of X-ray phase analysis, but it was detected in SEM images (see below).

It should be noted that the surface roughness ($R_a$) for the obtained eutectic samples without Ce dopants is in the range of 10.5-12.25 nm. For samples of Ce-doped eutectics, this is 11.24-14.48 nm under the same polishing conditions. Higher surface roughness values for doped samples are most likely due to differences in the mechanical characteristics of the $Al_2O_3$/YAG:Ce eutectic compared to $Al_2O_3$/YAG [20].

As mentioned earlier, the microstructure of the eutectic depends on both the pulling rate [18] and the dopant concentration [12]. Therefore, to establish the nature of the dependence of $\lambda_{eut}$ on the pulling speed, $Al_2O_3$/YAG eutectic was first obtained without adding the dopant. Over the entire considered range of pulling rates from 5 mm/h to 50 mm/h, the eutectic structure looked like a "Chinese script" (Fig. 3a-c). In the SEM images, the YAG phase appears lighter.

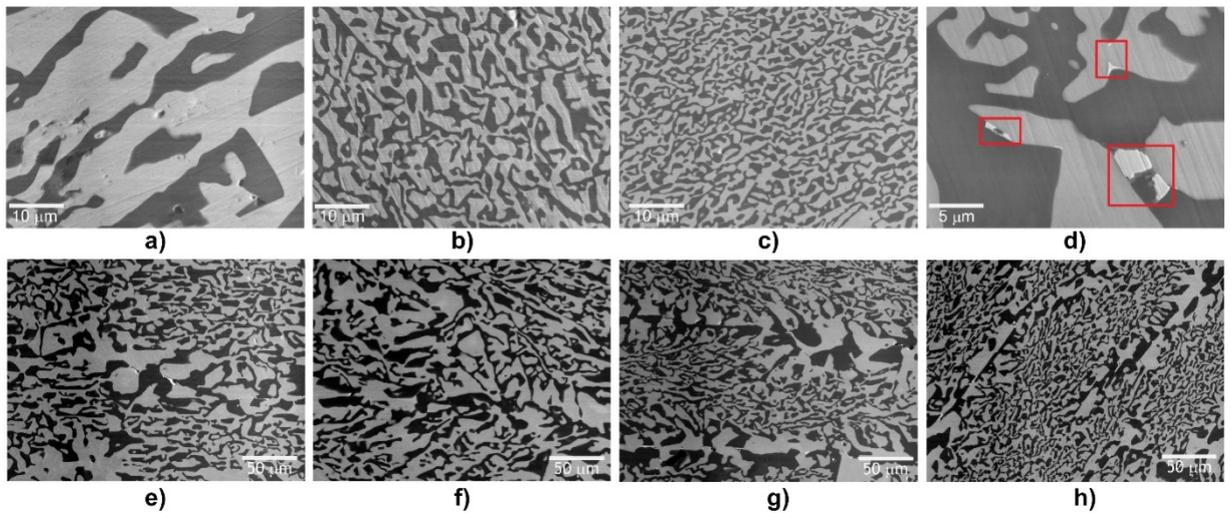

Fig. 3. SEM images of microstructure of $Al_2O_3$/YAG eutectic samples obtained at different pulling speeds and cut perpendicular to the pulling direction: $a$ – 5 mm/h; $b$ – 30 mm/h; $c$ – 50 mm/h; $d$ – $Al_2O_3$/YAG:$Ce^{3+}$ with a cerium concentration of 0.5 at.% ($V$ = 30 mm/h); the location of the Ce-enriched phase is highlighted by frames; $e, f$ - $Al_2O_3$/YAG:$Ce^{3+}$ with a cerium concentration of 0.25 at.% ($V$ = 30 mm/h), cut perpendicularly and along the pulling direction, respectively; $g, h$ – $Al_2O_3$/YAG:$Ce^{3+}$ with a cerium concentration of 0.5 at.% ($V$ = 30 mm/h), cut perpendicularly and along the pulling direction, respectively.

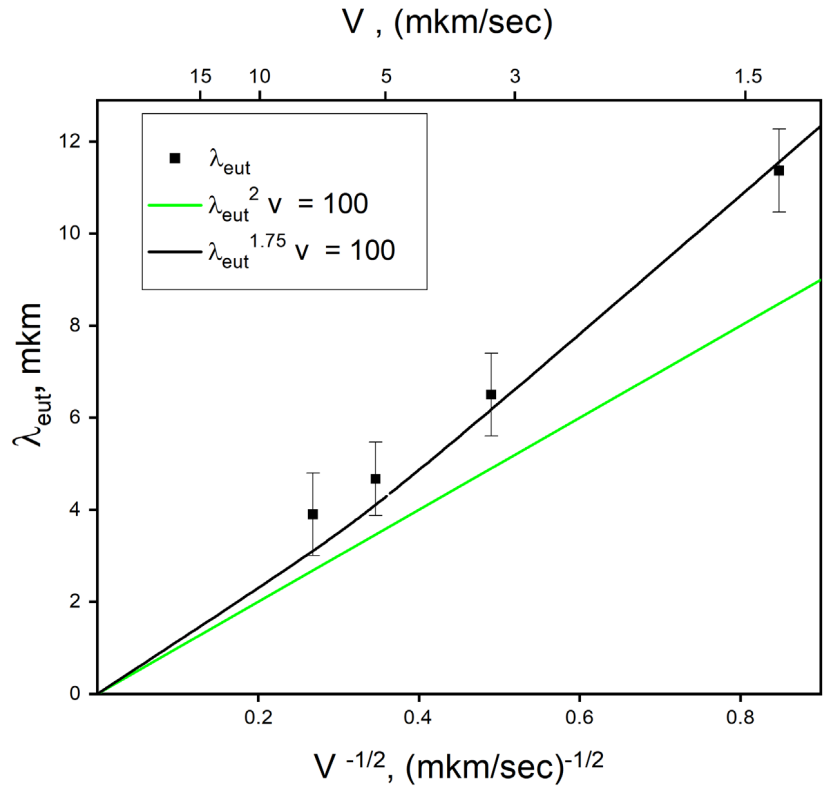

Fig. 4. Dependence of the characteristic eutectic interval ($\lambda_{eut}$) on the pulling speed for $Al_2O_3$/YAG eutectics. A straight line is a theoretical dependence corresponding to the formula $\lambda^2 \times V = const$ [14].

In the resulting $Al_2O_3$/YAG eutectic samples, which are cut perpendicular to the pulling direction, domain elongation along a certain direction is observed (Fig. 3e, g), which is explained by the presence of radial temperature gradients along the liquid–solid boundary.

Since in eutectic systems with a high probability of faceted growth (so-called "faceted" eutectics), which also includes the $Al_2O_3$/YAG system, the stability of the crystallization front may be disturbed, it can be assumed that in the formula $\lambda_{eut}^n \times V = const$ the exponent can deviate from an integer value ($n \neq 2$). As shown in Fig. 4, the $\lambda_{eut}$ values obtained for the $Al_2O_3$/YAG eutectic correspond to the dependence $\lambda_{eut}^{1.75} \times V = const$ (continuous curve in Fig. 4). This deviation may be related to the branching of domains during the formation of eutectics with a microstructure of this type, which is also indicated in [16].

At pulling speeds of 5 and 15 mm/h, in addition to a rather large value of the $\lambda_{eut}$ parameter, there is also a larger scatter in the sizes of domains of individual phases (especially at a speed of 5 mm/h), and at 50 mm/h, an increase in the number of micro defects (pores and cracks) is observed. Additionally, in certain areas of the resulting ingot, nucleation of a cellular structure occurs, which appears when the critical value of supercooling is exceeded according to the well-known criterion [27]. Based on this criterion, with a constant value of the temperature gradient at the interphase boundary and the amount of foreign impurities in the feedstock, the transition to cellular growth due to a violation of the stability of the microstructure morphology occurs when the critical value of the curing speed is exceeded, which in our case is 50 mm/h. Therefore, in the considered range of pulling speeds, a more homogeneous and stable microstructure of the eutectic (the smallest scatter of eutectic distance values, the minimum number of micropores and cracks) was obtained at a pulling speed of 30 mm/hour. To obtain a more accurate picture of the dependence of $\lambda_{eut}$ on growth parameters, additional more detailed studies are needed.

In the resulting doped $Al_2O_3/YAG:Ce^{3+}$ eutectic, a well-defined domain texture is observed both along and perpendicular to the pulling direction (Fig. 3 e-h). Moreover, the direction of the texture in different places of the ingot can change, since, as shown in [12], the phases that form the eutectic twist and branch during solidification. Additionally, in the SEM images of $Al_2O_3/YAG:Ce^{3+}$ eutectic samples, a "cellular structure" is observed - distribution into areas with a fine (segment CD in Fig. 5c) and coarse (segment AB in Fig. 5c) structure. Such a structure arises as a result of the transition from normal to cellular growth when the critical value of supercooling is exceeded according to the concentration supercooling criterion, the formation of a non-flat crystallization front, uneven distribution of cerium ions, and variations in the local crystallization rate [27].

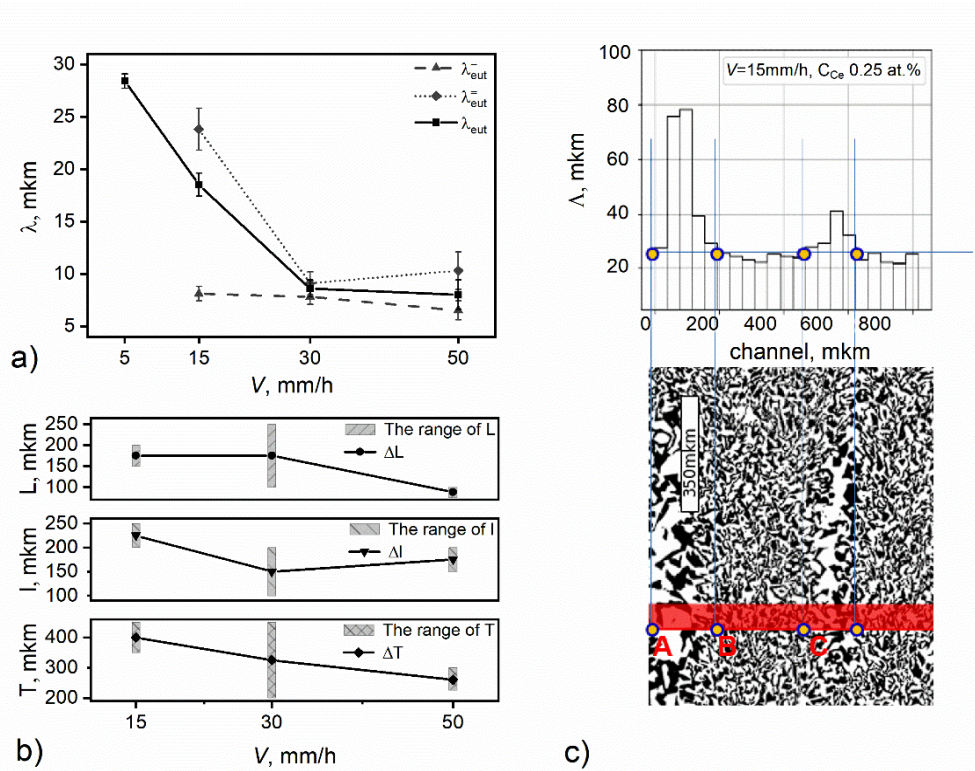

Fig. 5. Calculated parameters of the eutectic $Al_2O_3/YAG:Ce^{3+}$ (Ce 0.25 at.%): *a* - dependence of parameters $\bar{\lambda}_{eut}$, $\bar{\bar{\lambda}}_{eut}$ and $\lambda_{eut}$ on $V$; *b* – the minimum and maximum dimensions of the coarse microstructure *L*, the fine microstructure *l*, and the period of the cellular microstructure *T* and the dependence of their average values on $V$; *c* - an example of calculating the dimensions of the coarse *L* (marked in the red area, limited by segment AB), fine *l* (marked in the red area, limited by segment BC) microstructures, the period of the cellular microstructure *T* (marked in red, the area limited by segment AC) and the values of $\bar{\lambda}_{eut}$, $\bar{\bar{\lambda}}_{eut}$.

From the above description of the morphology of the doped Al$_2$O$_3$/YAG:Ce eutectic, it follows that to characterize it, there is a need to determine not only $\lambda_{eut}$ but also other parameters that can be used to more accurately describe the cellular structure: the sizes of the coarse (*L*) and fine (*l*) zones of the microstructure, period of the cellular structure (*T*), and eutectic distances in the fine ($\bar{\lambda}_{eut}$) and coarse ($\bar{\bar{\lambda}}_{eut}$) microstructures.

It can be accepted that

$$T \approx L + l; \qquad (2)$$

$$\lambda_{eut} \approx \frac{\bar{\lambda}_{eut} \times l}{T} + \frac{\bar{\bar{\lambda}}_{eut} \times L}{T}. \qquad (3)$$

Since the heterogeneity of the microstructure over the volume of the eutectic can affect its mechanical, optical, lighting, and other characteristics of the material, it is necessary to study the influence of Ce concentration and other crystallization conditions on these parameters. The parameters of the cellular structure of the eutectic are analysed both visually from SEM images and using various calculation methods. In many cases, it is quite difficult to clearly determine the boundary between zones visually, which increases the measurement error and can affect the final conclusion.

The modified chord method used in this work during step-by-step scanning of individual sections of the SEM image (Fig. 5c) allowed us to collect a sufficient amount of data to more accurately determine the boundary between the coarse and fine structure, especially in places with a blurred boundary. Additionally, in parallel, scanning profiles were compared with the SEM image in places with a clear boundary between the zones to verify the results obtained. The obtained research results are presented in Table 1 and Fig. 5. By analysing them, one can trace the change in the periodicity of the cellular structure and $\lambda_{eut}$ (for areas with fine and rough microstructure) with a change in the pulling speed and concentration of impurities.

Table 1. Parameters of the Al$_2$O$_3$/YAG:Ce$^{3+}$ eutectic structure (pulling speed 15 mm/h).

| Initial Ce concentration, at.% | Cellular structure period, $T$, μm | Coarse microstructure size, $L$, μm | Fine microstructure size, $l$, μm | Eutectic distance in coarse microstructure, $\bar{\bar{\lambda}}_{eut}$, μm | Eutectic distance in fine microstructure, $\bar{\lambda}_{eut}$, μm | Characteristic eutectic distance $\lambda_{eut}$, μm |
|---|---|---|---|---|---|---|
| 0 | No cellular structure | | | | | 6.5±0.9 |
| 0.25 | 350-450 | 150-200 | 200-250 | 23.8±2 | 8.1±0.7 | 18.5±1.1 |
| 1 | 300-550 | 175-275 | 250-375 | 26.8±1 | 14.7±0.6 | 22.9±0.8 |

The general tendency of changes in the period of the cellular structure, the sizes of the zones of the fine and coarse microstructure depending on the concentration of the dopant and the speed of crystallization is known [14], but according to the results of this work, this process is not completely uniform, and at a certain crystallization speed depending on the concentration of the dopant, there is a change in the nature of the restructuring of the structure, which is reflected by the inflection on the dependence of various parameters of the macro- and microstructure (Fig. 5 a, b). This indicates the existence of two stages in the formation of the colony structure of the doped eutectic. At the initial stage, an increase in the pulling speed to 15 mm/h (at a cerium concentration in the charge of 0.25 at.%) first causes the appearance of a cellular structure; then (with an increase in speed to 30 mm/h), its period decreases due to a decrease in the size of zones of fine microstructure (the size of zones of coarse microstructure practically does not change), and the characteristic eutectic distances $\lambda_{eut}^=$ decrease mainly in the zones with a coarse microstructure. However, when the speed, which depends on the dopant concentration, exceeds a critical value $V^*$ (over 30 mm/h for a concentration of 0.25 at.% (Fig. 5 a, b)), this trend changes to the opposite: the size of zones with a fine microstructure slowly increases or remains almost unchanged, and the size of zones with a coarse microstructure, on the contrary, quickly decreases, while in these zones, the parameter $\lambda_{eut}^-$ decreases, and the parameter $\lambda_{eut}^=$ increases. The period of the cellular structure decreases in both stages (see Fig. 5a, b). The inhomogeneity (scatter) of both the zone sizes and periodicity of the cellular structure was found in samples cut from different places in the ingot and may be associated with fluctuations in temperature and Ce concentration along the interface. As the pulling speed increases, the amount of dopant increases at the interface boundary since it does not have time to be pushed into the melt; this leads to the emergence of radial gradients of concentration and mass transfer of cerium along a nonplanar crystallization front. As a result, the dopant accumulates in the recesses between the cells and locally slows down the crystallization rate, which contributes to the formation of large-sized domains of individual phases in areas with a coarse microstructure.

The most correct way to characterize the heterogeneity of the microstructure of the doped eutectic $Al_2O_3$/YAG:$Ce^{3+}$ is the use of the parameters $\lambda_{eut}^-$ and $\lambda_{eut}^=$: a more stable and uniform microstructure of the eutectic will have a minimal difference between $\lambda_{eut}^-$ and $\lambda_{eut}^=$. As a result of experiments and studies of eutectics with different concentrations of cerium, it was found that the minimum heterogeneity of the microstructure along the plane of the sample and the lowest value of $\lambda_{eut}$ can be obtained under certain growth conditions. The smallest value of $\lambda_{eut}$ and the minimum difference between $\lambda_{eut}^-$ and $\lambda_{eut}^=$ were achieved in the samples at the critical speed $V^*$. For example, for a sample with a Ce concentration in the raw material of 0.25 at.% at a pulling speed of 30 mm/h, the value $\lambda_{eut}$ = 8.6±0.8 μm and the difference in 2.3±1.1 μm between $\lambda_{eut}^-$ and $\lambda_{eut}^=$ (which is ≈ 25%) were obtained (see Fig. 5 a, b).

An increase in the concentration of cerium leads to an increase in all parameters of the structure (macro- and micro) and an increase in the scatter (heterogeneity) of the macrostructure (*l*, *L* and *T*) (Table 1).

Thus, local changes in the morphology of the doped eutectic (the size of the zones of coarse and fine structure, the parameters $\lambda_{eut}^-$ and $\lambda_{eut}^=$), the period of the cellular structure and the parameter $\lambda_{eut}$ depend on both factors: the dopant concentration in the melt, its distribution along the crystallization front, and the growth rate.

The dependence of the values of the average eutectic distances $\lambda_{eut}^{Al_2O_3}$ and $\lambda_{eut}^{YAG}$, measured over the entire plane of the SEM image of $Al_2O_3$/YAG:$Ce^{3+}$ eutectic samples, on the Ce concentration was studied. It has been established that the curves of $\lambda_{eut}^{Al_2O_3}$ and $\lambda_{eut}^{YAG}$ versus Ce concentration, obtained by measurements in a certain direction, have an inflection at $C_{Ce} \approx 0.25$ at.%. This effect is most pronounced when scanning samples in the direction with the maximum domain size (Fig. 6).

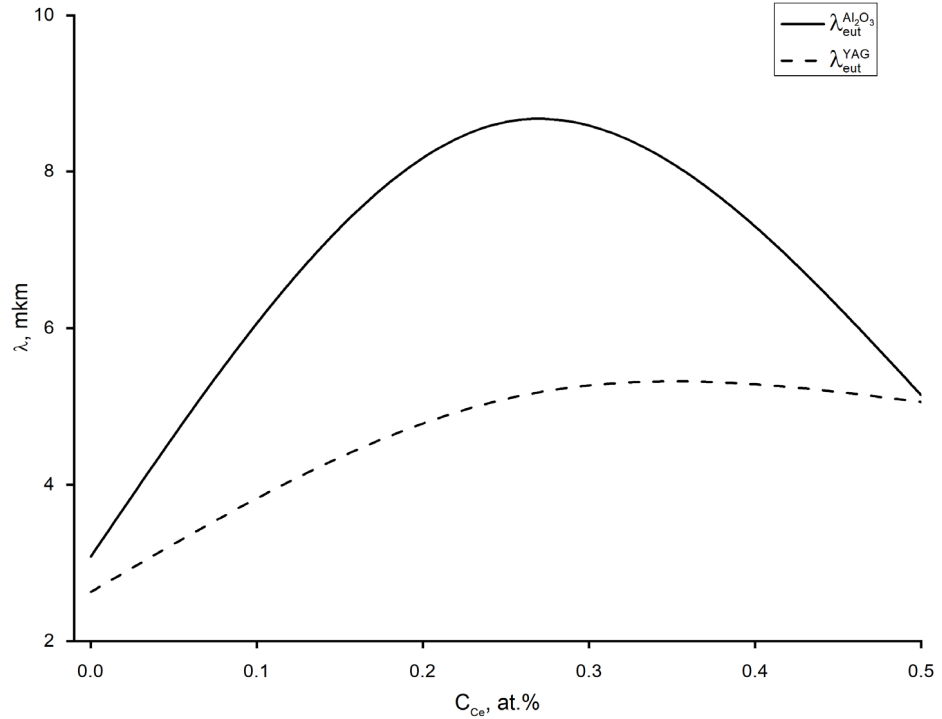

Fig. 6. Dependence of the values $\lambda_{eut}^{Al_2O_3}$ and $\lambda_{eut}^{YAG}$ on the concentration of Ce in the initial charge when measured in the direction with the maximum domain size.

Thus, it can be seen that there is a relationship between the sizes of $\lambda_{eut}^{Al_2O_3}$ and $\lambda_{eut}^{YAG}$ and the amount of dopant at the interface boundary. The accumulation of Ce impurities in front of the phase interface boundary leads to a gradual increase in the values of $\lambda_{eut}^{Al_2O_3}$ and $\lambda_{eut}^{YAG}$, while a decrease in the Ce concentration (for example, due to capture by the eutectic composite during solidification) will lead to a decrease in $\lambda_{eut}^{Al_2O_3}$ and $\lambda_{eut}^{YAG}$, which was observed.

As follows from these results, the concentration $C_{Ce} \approx 0.25$ at.% is critical, after which the active formation of a Ce-enriched phase is possible. An increase in the amount of the third phase, enriched in Ce, can be observed in the SEM images of samples with a Ce concentration of more than 0.25 at.% (Fig. 3d). The third phase (highlighted in red frames) is usually located in the area of coarse structure. Based on the analysis of phase diagrams of the $Al_2O_3$-$Ce_2O_3$ and $Al_2O_3$-$CeO_2$ systems, as well as data on the mutual solubility of $CeO_2$ in $Y_2O_3$ [28] and literature sources [14, 29, 30], we assume that this may be the $CeAl_{11}O_{18}$ or $CeAlO_3$ phase. The cerium-rich third phase can arise when the dopant concentration exceeds the solubility limit in the eutectic melt and does not have time to be removed from the interface boundary by diffusion or convection of the melt.

## 4. Conclusions

Thus, $Al_2O_3$/YAG and $Al_2O_3$/YAG:$Ce^{3+}$ (up to 1 at.% Ce) eutectics with the "Chinese script" type microstructure were obtained by the HDC method in a carbon-containing reducing atmosphere based on Ar in the range of pulling speeds of 5-50 mm/h. The possibility of obtaining an $Al_2O_3$/YAG eutectic with a stable, homogeneous microstructure and a characteristic eutectic distance $\lambda_{eut}$ of approximately 4 μm at a pulling rate of 30 mm/h and a temperature gradient of 45 K/cm has been proven.

To characterize the microstructure, a modified method of chords was used, which made it possible to identify the features of changes in the parameters of the eutectic microstructure depending on the growth conditions.

For $Al_2O_3$/YAG eutectics, a deviation from the quadratic dependence $\lambda^2_{eut} \times V = const$ obtained for regular eutectics according to the Jackson-Hunt model was established. In the range of eutectic pulling speeds of 5-50 mm/h, the relation $\lambda_{eut}^{1.75} \times V = const$ is most likely to be fulfilled. This corresponds to the fractal growth of eutectics and may be due to the lateral growth of the domains of individual phases and the nonlinear nature of the diffusion of components.

Based on the dependence of the microstructure parameters of the doped $Al_2O_3$/YAG:$Ce^{3+}$ eutectic on the Ce content and the growth rate, a two-stage process of development of the morphological instability of the crystallization front and the microstructure of the eutectic was established. When the growth rate increases, the size of zones with a fine microstructure decreases, while the size of zones with a coarse microstructure is almost unchanged. At the same time, the characteristic eutectic distances $\lambda_{eut}^-$ and $\lambda_{eut}^=$ decrease in both zones, and the period of the cellular structure also decreases. When the growth rate exceeds a certain value, which depends on the cerium concentration, the sizes of zones with a fine microstructure begin to increase, and the sizes of zones with a coarse microstructure begin to decrease. The eutectic distance of the fine microstructure $\lambda_{eut}^-$ continues to decrease, and the distance in the coarse microstructure $\lambda_{eut}^=$ begins to increase. At the same time, the period of the cellular structure gradually decreases. Such changes in eutectic morphology parameters indicate a two-stage process of development of morphological instability of the crystallization front and eutectic microstructure.

The predominant dependence of local changes in the morphology of the eutectic (sizes of zones of coarse and fine structure, parameters $\lambda_{eut}^-$ and $\lambda_{eut}^=$) on crystallization conditions was established. The change in the period of the cellular structure and the parameter $\lambda_{eut}$ depends on both factors – the concentration of Ce in the melt and the growth rate. The parameters of the $Al_2O_3$/YAG:$Ce^{3+}$ eutectic structure increase with increasing dopant concentration. A stable and most homogeneous eutectic microstructure was obtained with the lowest value of $\lambda_{eut}$ (8.6±0.8 μm) and the minimum difference between $\lambda_{eut}^-$ and $\lambda_{eut}^=$ (2.3±1.1 μm) at a pulling speed of 30 mm/h and Ce concentration of 0.25 at.% in the raw material.

The Ce concentration ≈0.25 at.% in the initial charge was determined as a threshold limit, after which the formation of a third phase enriched in Ce actively occurs under the given conditions for obtaining the eutectic.